# DECREASE OF TRANSIT-TIME BROADENING OF SPECTRAL RESONANCES BY OPTICAL PUMPING OF ATOMS

Azad Ch. Izmailov

Institute of Physics, Azerbaijan National Academy of Sciences, Javid av. 33, Baku, Az-1143, AZERBAIJAN

*e-mail*: azizm57@rambler.ru

*ABSTRACT*. **We propose the new method of essential decrease of transit-time broadening of Doppler-free absorption resonances on transitions between long-lived quantum levels of atoms (or molecules) of a rarefied gas medium. This method is based on preliminary optical pumping of the ground term of atoms (molecules) by additional radiation which is spatially separated from the recording light beam. The proposed method may be applied in atomic (molecular) spectroscopy of ultra high resolution and also in frequency and time standards.**

## 1. INTRODUCTION

Development of effective methods of essential decrease not only Doppler but also transit-time broadening of recorded spectral resonances on optical transitions of atoms (molecules) in sufficiently rarefied gas media is very important for spectroscopy of ultra high resolution and optimization of time and frequency standards [1,2]. Indeed, in many experiments in laser spectroscopy, the interaction time of atoms (molecules) with radiation fields is small compared with spontaneous lifetimes of quantum levels of resonance transitions. Then, in the absence of a Doppler broadening, linewidths of corresponding recorded spectral resonances are determined, mainly, by effective transit times of atoms (molecules) through the recording light beam. By now, two following methods of decrease of such a transit-time broadening are elaborated [1,2]: one may either enlarge the laser beam diameter, or one may slow down of atoms (or molecules) by corresponding "cooling" techniques.

In the present work, we propose the new method to achieve this goal, which is based on preliminary optical pumping of the ground quantum term of atoms (or molecules) by the additional radiation which is spatially separated from the recording light beam. This method may be applied both independently and along with the noted known methods for further decrease of transit-time broadening of recorded spectral resonances.

Indeed, let us consider sufficiently simple experimental scheme in Fig.1, which presents the cross section of the recording light radiation with the diameter *d* and the coaxial cross section of the beam of broadband optical pumping. It is considered that the central radiation in Fig.1 records the spectral resonance on a transition *a→c* from sublevel *a* of the ground atomic term to the long-lived (in particular metastable) excited quantum state *c*. Such ultra narrow spectral resonances, for example, may take place in Doppler-free multiphoton spectroscopy [1]. At the same time, the coaxial broadband optical pumping (Fig.1) is realized on a non-cyclic transition *a→b* from the same lower level *a* to the excited state *b* (Fig.2). In experiments we may directly distinguish a contribution of optically pumped atoms to the analyzed resonance on the transition *a→c*. This is achieved by recording of the difference between signals of absorption (or dispersion) of the central light beam, which are subsequently detected with and without the spatially separated optical pumping (Fig.1). We will show that characteristic transverse components of velocities of optically pumped atoms, reaching the recording light beam, may be substantially reduce at a lowering of the pumping rate. Thus, essential decrease of the transit-time broadening of the spectral resonance on analyzed transition *a→c* may be achieved, because of recording of such a resonance directly on these comparatively slow moving pumped atoms.

## 2. BASIC RELATIONSHIPS

We consider a sufficiently rarefied gas medium, where the free path of atoms (molecules) substantially exceeds the radius $r_2$ of the beam of broadband optical pumping (Fig.1). The pump intensity on the transition *a→b* is assumed to be so low that population of excited state *b* is negligibly small relative to that of initial level *a* (Fig.2). Then, from known balance relations [1] we receive the following equation for the population $\rho_a(\mathbf{R}, \mathbf{v})$ of atoms on the long-lived level *a* with the velocity **v** and the coordinate vector **R**:

$$\mathbf{v}\frac{\partial \rho_a}{\partial \mathbf{R}} = -W(\mathbf{R})\rho_a, \qquad (1)$$

where $W(\mathbf{R}) = \epsilon_{ab}(\mathbf{R})(1 - B_{ba})$ is the rate of the broadband optical pumping, $\epsilon_{ab}(\mathbf{R})$ is the excitation probability of atoms by the pumping radiation on the non-cyclic transition *a→b*, the constant $B_{ba}$ is the probability of the subsequent radiative decay through the channel $b \to a$ ($B_{ba} < 1$). It is assumed, that atoms have equilibrium distributions both on their velocities and populations of quantum levels before interaction with the pumping radiation. Under the above conditions, population $\Delta\rho_a(\mathbf{R}, \mathbf{v})$ of the optically pumped atoms in the lower level *a* is described by expression:

$$\Delta\rho_a(\mathbf{R},\mathbf{v}) = n_a F(\mathbf{v}) - \rho_a(\mathbf{R},\mathbf{v}), \qquad (2)$$

where $n_a$ is equilibrium density of atoms in the level $a$ with the Maxwell distribution $F(\mathbf{v})$ on atomic velocities $\mathbf{v}$. We will consider the cylindrically symmetric distribution of the optical pumping rate $W(r)$ (1) on the distance $r$ from the central axis of recording light beam (Fig.1). Then, from Eq.(1), we receive the following expression for the population $\Delta\rho_a$ (2) of optically pumped atoms, reaching the region of sufficiently narrow recording light beam with the diameter $d \ll r_1$ (Fig.1):

$$\Delta\rho_a(v_r, s) = n_a F_r(v_r)\left[1 - exp\left(-\frac{su}{v_r}\right)\right], \qquad (3)$$

where the dimensionless parameter $s$ characterizes efficiency of the optical pumping and has the following form:

$$s = \left(\frac{1}{u}\right)\int_{r_1}^{r_2} W(r)dr, \qquad (4)$$

and $F_r(v_r)$ is the Maxwell distribution on radial (transverse) component $v_r$ of atomic velocity:

$$F_r(v_r) = 2v_r u^{-2} exp(-v_r^2 u^{-2}), \qquad (5)$$

with the most probable atomic speed $u$ in the gas.

Let us consider, for example, Doppler-free absorption spectral line, whose profile has following Lorentzian shape for atoms (molecules) with transverse velocity component $v_r$:

$$L(\delta, v_r) = \frac{\gamma(v_r)}{[\gamma(v_r)^2 + \delta^2]}, \qquad (6)$$

where $\delta$ is the frequency detuning of the recording light beam (Fig.1) from the center of the analyzed quantum transition $a \rightarrow c$ (Fig.2). The characteristic width $\gamma(v_r)$ of the given line (6) is the sum of two following terms:

$$\gamma(v_r) = \gamma_0 + \frac{v_r}{\langle v_r \rangle}\gamma_{tr}, \qquad (7)$$

where $\gamma_0$ is the radiative width of this line, which is determined by the lifetime of the excited level $c$, and the value $\gamma_{tr}$ (in the second phenomenological term) is the transit-time relaxation rate in the absence of the optical pumping at the average transverse component of atomic velocity $\langle v_r \rangle = \int_0^\infty v_r F_r(v_r)dv_r \approx 0.886u$.

From Eqs. (3), (6) we receive the integral absorption profile directly for optically pumped atoms with all possible velocity components $v_r$:

$$\Delta J(\delta, s) = \int_0^\infty \Delta\rho_a(v_r, s) L(\delta, v_r) dv_r .  \qquad (8)$$

In the absence of the optical pumping, corresponding profile for nonpumped atoms has the form:

$$J_0(\delta) = n_a \int_0^\infty F_r(v_r) L(\delta, v_r) dv_r . \qquad (9)$$

### 3. DISCUSSION OF RESULTS AND CONCLUSION

Fig.3 presents dependences of the population $\Delta\rho_a$ (3) of optically pumped atoms on their transverse velocity component $v_r$ at different intensities of the optical pumping. For sufficiently large values of the parameter $s \gg 1$ (4), effective pumping practically all atoms occurs and then corresponding dependence $\Delta\rho_a(v_r, s)/n_a$ (curve 4 in Fig.3), in fact, coincides with the equilibrium velocity distribution $F_r(v_r)$ (5). At the same time, we see in Fig.3 that lowering of this pumping parameter $s$ causes not only decrease in the number of optically pumped atoms but also to effective shift of the distribution $\Delta\rho_a(v_r, s)$ to smaller values of their velocity components $v_r$. Thus, it is possible significantly to reduce transit-time broadening of spectral resonances recorded directly on given pumped atoms.

Further we will consider situation when characteristic rate $\gamma_{tr}$ of transit-time relaxation of atoms in the absence of the optical pumping is many times greater than the radiative relaxation rate $\gamma_0$ (7) of analyzed spectral resonance. Fig.4 presents spectral absorption profiles $\Delta J(\delta, s)$ (8), which are recorded directly on optically pumped atoms at various pumping parameters $s$ (4). Lowering of the pumping intensity leads to decrease both amplitude $A$ and width $W$ of such spectral resonances. Indeed, Fig.5 presents corresponding values $A(s) = \Delta J(\delta = 0, s)$ and $W$ (on the half-height) versus the pumping parameter $s$ (4). With the growth of the parameter $s$ in the region $s>1$, these quantities $A$ and $W$ asymptotically approach to values $A_0$ and $W_0$ corresponding to the spectral profile $J_0(\delta)$ (9) for atoms in the absence of the optical pumping. We see in Fig.5b that essential decrease of transit-time broadening of the recorded spectral resonance may be achieved at lowering of the optical pumping. For example, according to Fig.5, at the pumping parameter $s=0.02$, the width $W$ of the resonance decreases about 5 times in comparison with the corresponding value $W_0$ obtained in the absence of the optical pumping, although amplitude $A$ of this resonance is about 9% of the maximum value $A_0 = J_0(\delta = 0)$ (9).

We have considered comparatively simple experimental scheme (Fig.1) with the cylindrical symmetry of light beams. However, in the similar manner, it is possible to analyze theoretically and then implement in practice another schemes with spatially separated beams of recording radiation and optical pumping for effective decrease of transit-time broadening of detected spectral resonances.

## References

1. *Demtroder W.* Laser Spectroscopy: Basic Concepts and Instrumentation. (Springer, 2003).
2. *Riehle F.* Frequency Standards-Basics and Applications. Berlin: Wiley-VCH, 2004.

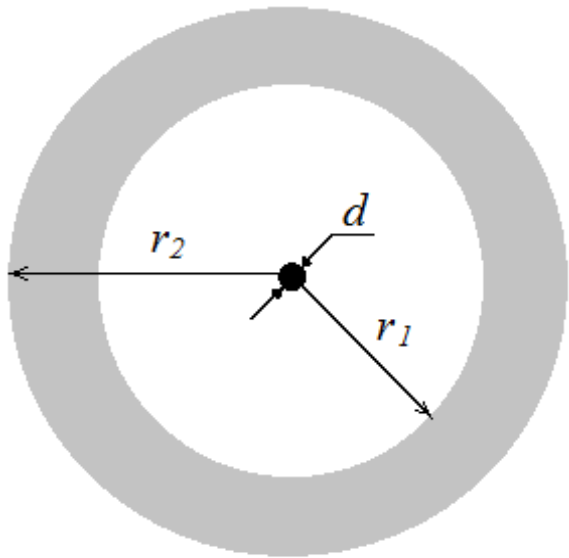

**Fig.1.** Scheme of cross section of the recording light beam with the diameter $d$ and the coaxial spatially separated optical pumping radiation limited by radiuses $r_2 > r_1$.

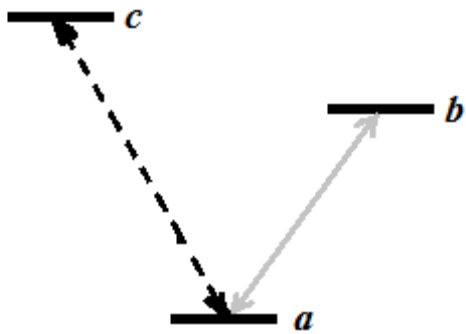

**Fig.2.** Diagram of atomic quantum levels and transitions, where *a→c* is the analyzed transition and *a→b* is the pumping transition from the ground level *a*.

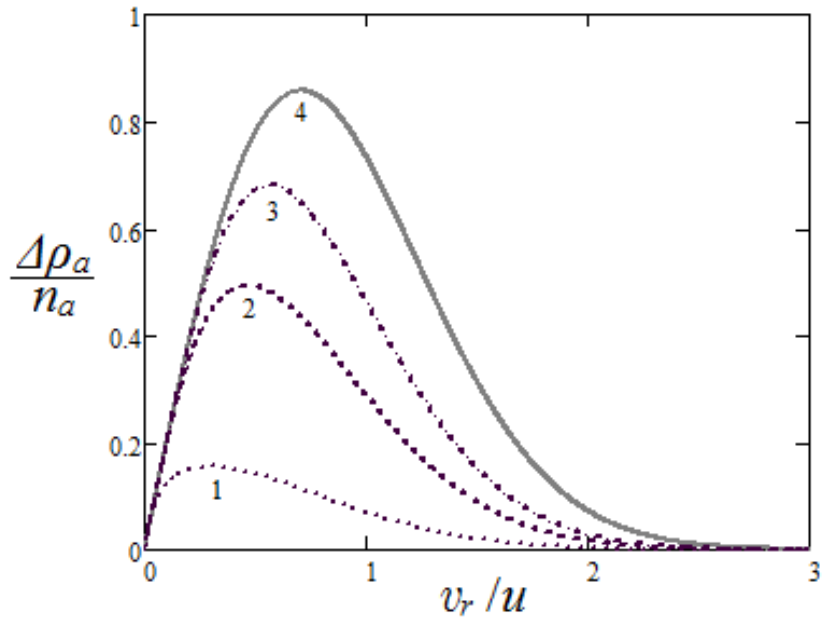

**Fig.3.** Distribution $\Delta\rho_a(v_r, s)$ of optically pumped atoms on the transverse component $v_r$ of their velocity at the pumping parameter $s$=0.1 (curve 1), 0.5 (2), 1 (3), and 10 (4).

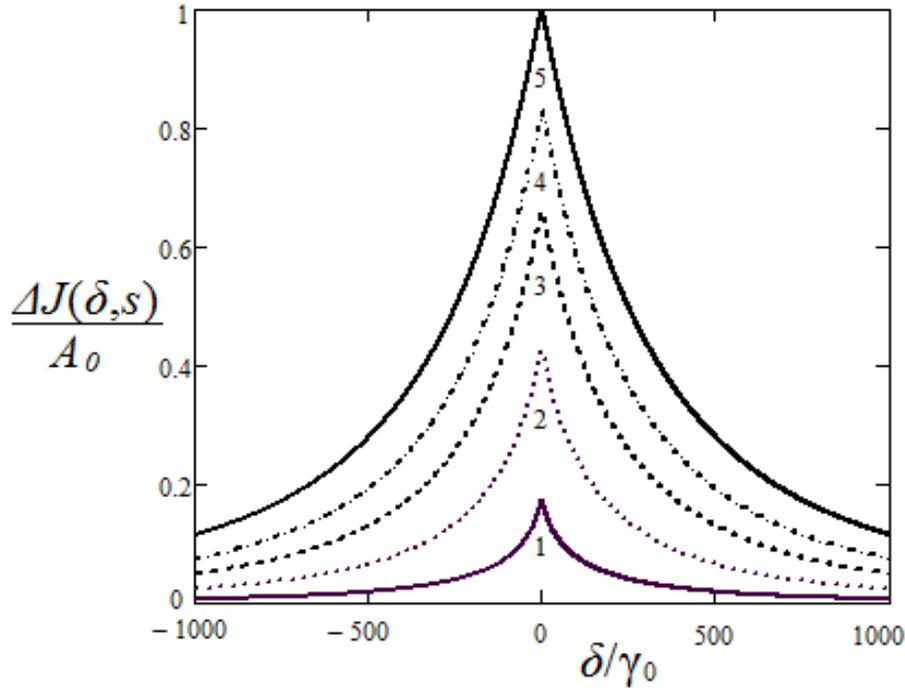

**Fig.4.** Spectral absorption resonance $\Delta J(\delta, s)$ of optically pumped atoms versus frequency detuning δ in units $A_0 = J_0(\delta = 0)$ (9), when $\gamma_0 = 2 * 10^{-3} \gamma_{tr}$ and the pumping parameter $s$=0.05 (curve 1), 0.2 (2), 0.5 (3), 1 (4) and 10 (4).

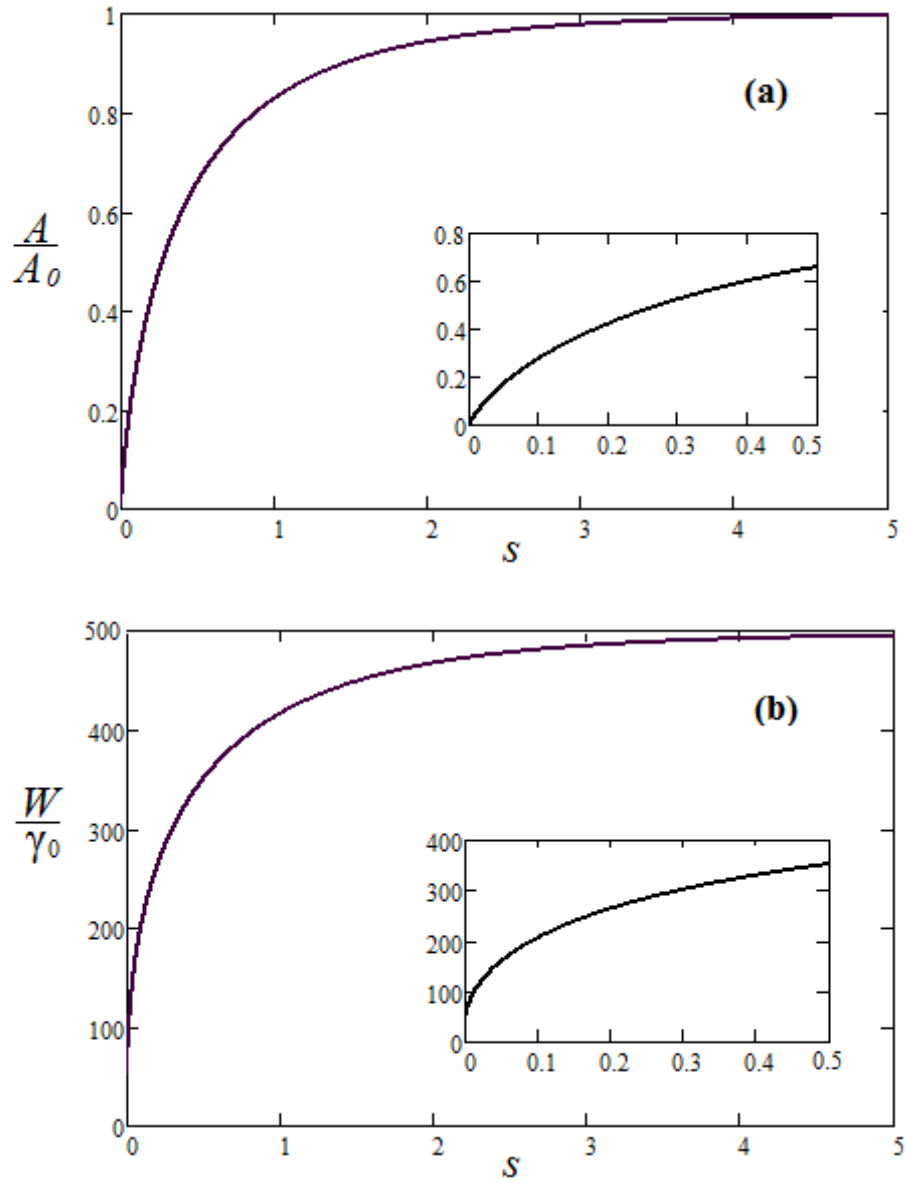

**Fig.5.** Amplitude *A* (a) and width *W* (b) of the spectral absorption resonance versus the pumping parameter *s*, when $\gamma_0 = 2 * 10^{-3}\gamma_{tr}$. Insets present corresponding dependences in the region of sufficiently small values *s*< 0.5.